# Density of vacuum energy for multidimensional model of Kazner with scalar field and cosmological birth of particles

Sergey V. Yakovlev
sergey.y@mail.ru

Moscow, April 23, 2011

In the work's considered density of vacuum energy and dynamic of scalar field in multidimensional theory with cosmological constant. Using method of N.N.Bogolubov coefficients, was gotten expression for influence of anisotropic metric to vacuum energy. Obtained the effective mass of massles scalar field, that depends on cosmological constant, and some general theoretical results.

Multidimensional anisotropic Kazner model allowed to use suitable form of metric with Kazner conditions on functions, included into metric. That metric for the case of three space-like surface at first time was used by Edvard Kazner in 1922, so people've got the possibility to investigate effects of anisotropy in Einstein equations. Using multidimensional world is interesting for considering anisotropic manifolds with isotropic submanifolds, and to make solving of tasks a little simpler, we usually take diagonal metric. An ordinary plus of Kazner metric is, that it describes flat space, that up to now is unexplainable fact in the Universe. Influence of anisotropy energy of space-time curvature is supposed could be higher that the density of momentum-energy of matter in such theory.

Exactly solvable task is considering the evolution of anisotropic Kazner Universe with cosmological term. So we could give it a try to consider dynamic of the fields, and for the most part scalar field, on the ground of evolution of anisotropic metric, with saving condition of prevailing of anisotropy energy, that is in Einstein equation we can neglect the tensor of energy-momentum of investigated field. Earlier was considered model with massless scalar field, and was showed that presence of such a field «generates» additional space dimensions [16]. That dimensions could be included into the metric by renormalizing of metric coefficients and cosmological constant. After that procedure metric's got the form of Kazner type, and satisfied to Kazner conditions. This task represents a private interest in investigation of effects in anisotropic multidimensional Kazner metric with the presence of matter.

Let's consider massive scalar field in Gaussian normal system on the ground of next anisotropic metric:

$$ds^2 = -dt^2 + e^{2\alpha(t)} e^{2\beta_i(t)} (dx^i)^2. \qquad (1)$$

The Klein-Gordon equation for the scalar field with the Lagrangian

$$L = -\frac{1}{8\pi}(\phi_{,\alpha}\phi^{,\alpha} + m^2\phi^2)$$

is

$$\frac{1}{(-g)^{1/2}}\left(\phi^{,\alpha}(-g)^{1/2}\right)_{,\alpha} - m^2\phi = 0. \qquad (2)$$

From here we could get the next equation:

$$\ddot{\phi} + n\dot{\alpha}\dot{\phi} - \phi_{,ii}e^{-2\alpha-2\beta_i} + m^2\phi = 0. \tag{3}$$

Supposed uniformity of space, we'll look for the decision of that equation in the form

$$\phi = \frac{1}{(2\pi)^{n/2}} e^{-n\alpha/2} e^{-i\vec{k}\vec{x}} \chi(t).$$

Using (3), we obtain the next expression,

$$\ddot{\chi} + (m^2 + \sum_i k_i^2 e^{-2\alpha(t)-2\beta_i(t)} - \frac{n}{2}\ddot{\alpha} - \frac{n^2}{4}\dot{\alpha}^2)\chi = 0. \tag{4}$$

Now let's recall concrete Einstein equations for metric type like (1):

$$\dot{\alpha}^2 = \frac{1}{n(n-1)}\left(2\Lambda + \sum_i \dot{\beta}_i^2\right) \tag{5}$$

$$\ddot{\beta}_j + n\dot{\alpha}\dot{\beta}_j = 0$$

Getting from here $\dot{\alpha}$, $\ddot{\alpha}$ and substitute it into (4), and remembering that [15]

$$\dot{\beta}_k = B_k e^{-n\alpha}, \quad B^2 \equiv \sum_{k=1}^{n} B_k^2, \quad B_k = const,$$

we obtain the final equation for the field $\chi(t)$:

$$\ddot{\chi} + (m^2 - \frac{\lambda^2}{4} + \sum_i k_i^2 e^{-2\alpha(t)-2\beta_i(t)} + \frac{nB^2}{4(n-1)} e^{-2n\alpha(t)})\chi = 0, \tag{6}$$

where

$$\lambda^2 = \frac{2n\Lambda}{n-1}.$$

So we see that here appears effective mass $m_{eff}^2 = m^2 - \frac{\lambda^2}{4}$, which depends on the sign and value of cosmological constant $\Lambda$.

Equation (6) because of the difficulty could be solved by approximate manners, for instance by using WKB method. The most optimal is method, suggested by Zeldovich, Starobinsky, and Birell, Davis [11], and Jeffreys. We would take it for the basis. For simplicity we'll be satisfied getting the evaluation in the first approximation.

Let's derive coefficients of N.N.Bogolubov [12], which connect decisions for «in» and «out» state of eq.(6), defining conditions of vacuum for the metric (1). Field $\chi(t)$ we could decompose using orthonormalized basis of decisions of eq.(6) in following manner:

$$\chi(t) = \sum_i \left[ a_i u_i(x) + a_i^+ u_i^*(x) \right] \quad (7)$$

Consider another set of decisions $\bar{u}_i(x)$. We can decompose the field $\chi(t)$ using set

$$\chi(t) = \sum_i \left[ \bar{a}_i \bar{u}_i(x) + \bar{a}_i^+ \bar{u}_i^*(x) \right].$$

That decomposition defines new vacuum condition $\bar{a}_j |\bar{0}\rangle = 0$, $\forall j$.

Basic functions could be expressed each through the other by N.N.Bogolubov conversion [12], [13]:

$$\bar{u}_j = \sum_i (\alpha_{ji} u_i + \beta_{ji} u_i^*)$$

$$u_i = \sum_j (\alpha_{ji}^* \bar{u}_j - \beta_{ji} \bar{u}_j^*),$$

Mathematical expectation of particles number operator $N = a^+ a$ in new vacuum $\langle \bar{0} | N | \bar{0} \rangle \neq 0$, but is equal

$$\langle \bar{0} | N | \bar{0} \rangle = \sum_j |\beta_{ji}|^2, \quad (8)$$

that is vacuum of modes $\bar{u}_j$ contains $\sum_j |\beta_{ij}|^2$ of particles of the mode $u_i$.

Let's calculate N.N.Bogolubov coefficients, connecting «in» and «out» states for metric (1) at different moments of time. «Out» state we will take when $t \to \infty$, and «in» – when some sufficiently big $t_0 \gg 1/\lambda$, of course $t > t_0$.

From equation (6) we see that while scale factor is increasing, we could neglect time-dependable terms and «out» state will be defined by expression that describing usual coherent oscillations of background surrounding in proper frame of reference:

$$\ddot{\chi} + (m^2 - \frac{\lambda^2}{4})\chi = 0 \quad (9)$$

That oscillating field we could represent as the set of free particles with mass $m^2 - \frac{\lambda^2}{4}$ in coherent state. Normalized positive-frequency decision of eq.(9) when $t \to \infty$ has the form

$$\chi^{out}(t) = (2\omega)^{-1/2} e^{-i\omega t};$$

scalar product for functions $\phi(x)$ is introduced in following way

$$(\phi_1, \phi_2) = -i \int \{\phi_1(x) \partial_t \phi_2^*(x) - [\partial_t \phi_1^*(x)] \phi_2(x)\} \sqrt{-g(x)} d^n x,$$

where $g(x)$ - determinant of metric tensor (1). Normalization condition for $\chi(t)$ is looked like this:

$$\chi\dot{\chi}^* - \chi^*\dot{\chi} = i. \qquad (10)$$

When mass $m$ is equal zero, equation (9) is getting the form

$$\ddot{\chi} - \frac{\lambda^2}{4}\chi = 0.$$

That is when $\Lambda < 0$ there are oscillations with frequency

$$\omega = \sqrt{\frac{n|\Lambda|}{2(n-1)}}.$$

In the first approach Bogolubov coefficient $\beta$, responsible for the difference of vacuum states «in» and «out», has the form [11]

$$\beta = \frac{i}{2\omega}\int_{t_0}^{\infty} e^{-2i\omega t'} Q(t')dt', \qquad (11)$$

where

$$Q(t) = -\sum_i k_i^2 e^{-2\alpha(t)-2\beta_i(t)} - \frac{nB^2}{4(n-1)} e^{-2n\alpha(t)}, \quad \omega^2 = m^2 - \frac{\lambda^2}{4}, \quad \omega^2 > 0.$$

We'll be interested $|\beta|^2$, describing distribution of particles in «in» vacuum state. And we'll take the matter for certainty with $\Lambda > 0$.

Right away we could see, that expression for the density of particles will diverge because of integrating over momentum at infinite limits, so we could restrict integration within maximum momentum $k_{\max} = K$. Lower limit of integrating for Bogolubov coefficient (start time) we've denoted $t_0$, that corresponds «in» state.

Exact expressions of functions for scale factors in considered multidimensional anisotropic model with $\Lambda > 0$ have the following form

$$\alpha(t) = \frac{1}{n}\ln\left(\frac{B}{\sqrt{2\Lambda}} sh\lambda t\right), \quad \beta_k = \frac{B_k}{B}\sqrt{\frac{n-1}{n}}\ln\left(\frac{B}{\sqrt{2\Lambda}} th(\lambda t/2)\right),$$

$$\lambda = \sqrt{\frac{2\Lambda n}{n-1}}, \quad \Lambda > 0; \qquad (12)$$

and additional condition, that following from initial dates and could be used, is $B = \sqrt{2\Lambda}$ [15]. Asymptotic form for $Q(t)$, when $t \to \infty$, is writing down as following:

$$Q(t) = -\lambda^2 \exp(-2\lambda t) - A^i k_i^2 \exp(-2\lambda t/n), \qquad (13)$$

where

$$A^i = 4^{1/n} \left( \frac{\sqrt{2\Lambda}}{B} \right)^{2(1/n+C_i)},$$

$$C_i = \frac{B_i}{B} \sqrt{\frac{n-1}{n}}.$$

Substituting expression (13) into (11), we get the next formula for Bogolubov coefficient, which is responsible for difference in «in» and «out» vacuums at the moment $t_0$:

$$\beta_k = -\frac{i}{2\omega} \int_{t_0}^{\infty} e^{-2i\omega t} (\lambda^2 \exp(-2\lambda t) + A^i k_i^2 \exp(-2\lambda t/n))dt, \qquad (14)$$

Here we imply, that $t_0 \gg 1/\lambda$. We have to note that for the moment in real cosmological model in the exponents in eq.(14) we've got degree closed to zero because of the smallness of $\lambda$, and given approximation in reality describes future of Kazner's Universe.

$|\beta|^2$ with remained main term of asymptotic for time is equal

$$|\beta_k|^2 = \frac{(A^i k_i^2)^2 \exp(-4\lambda t_0/n)}{16(m^2 - \lambda^2/4)(m^2 - \lambda^2/4 + \lambda^2/n^2)}. \qquad (15)$$

In «out» area density of particle numbers in the unit of proper volume is equal

$$n = \frac{1}{\left(2\pi e^{\alpha(t)}\right)^n} \int |\beta_k|^2 d^n k, \qquad (16)$$

or

$$n \approx \frac{1}{8(2\pi)^n} \frac{\sqrt{2\Lambda}}{B} \frac{\exp(-(1+4/n)\lambda t) Z_k}{(m^2 - \lambda^2/4)(m^2 - \lambda^2/4 + \lambda^2/n^2)};$$

density of energy is equal

$$\rho = \frac{1}{\left(2\pi e^{\alpha(t)}\right)^n} \int |\beta_k|^2 \omega_0 d^n k, \qquad (17)$$

or approximately

$$\rho \approx \frac{1}{8(2\pi)^n} \frac{\sqrt{2\Lambda}}{B} \frac{\exp(-(1+4/n)\lambda t) D_k}{(m^2 - \lambda^2/4)(m^2 - \lambda^2/4 + \lambda^2/n^2)},$$

where $Z_k, D_k$ are constants, denoted offcut integrals over momentum in eq.(16) and (17) respectively.

Calculated value of the density of vacuum energy $\rho$ for the field $\phi$ should be more less than effective density of energy of anisotropy $\rho_{anis}$:

$$\rho \ll \frac{1}{16\pi} \sum_i \dot{\beta}_i^2 = \rho_{anis}, \qquad (18)$$

(here functions $\beta_i$ describe anisotropic metric (1), but not spectrum of new vacuum); first equation (5) is valid cause of it. From here we could get some restrictions for initial parameters of the model and describing time interval:

$$\frac{\sqrt{2\Lambda} D_k}{(2\pi)^n (m^2 - \lambda^2/4)(m^2 - \lambda^2/4 + \lambda^2/n^2)} \exp(-4\lambda t/n) \ll \frac{B^3}{2\pi},$$

that is started «velocities» of «big bang» have to be sufficiently big.
In other case will be observed conversed effect of influence of vacuum energy density to the dynamic of multidimensional Kazner Universe, that goes out of our model.

Let's consider behavior of the $|\beta_k|^2$ in eq.(15), assuming $n \gg 1$. We get

$$|\beta_k|^2 = \frac{k^4}{16(m^2 - \Lambda/2)^2},$$

where $k^2 = \sum k_i^2$. Number of arised particles is not depended on time, though here arises difficulty with integrating over momentum in (17), when we want to calculate density of energy in case $n \to \infty$.

All calculations were made in proper frame of reference, that is observer is moving along with expanding cosmological «liquid». Usually is introducing conformal metric. In our case physical sense of obtained results is obvious – they also relate to proper frame of reference, moving along with galaxies.

There should be presented restriction for probability of birth of high energy particles, and it could be explained by boundedness of anisotropy energy, which has gravitational character. So offcut of integral in (16), (17) is correct.

Given model could also illustrate effect concerned restriction for mass in eq.(6). If $\Lambda$ is sufficiently big, square of frequency become negative, that could be interpreted as instability of vacuum state, so it could possibly be the spontaneous break of symmetry if introduced additional term into the potential is in charge of interaction.

Efficiency of calculating for density of vacuum energy in the model could be improved by using alternative approach, demanded the straight calculation of energy-momentum tensor. In that manner could be obtained straight contributions into energy-momentum, conditioned by anisotropy of space-time in the model.

Energy-momentum tensor of scalar field for quadratic Lagrangian has the form

$$T_{\mu\nu} = \frac{1}{4\pi}\left(\phi_{,\mu}\phi_{,\nu} - \frac{1}{2} g_{\mu\nu}(\phi_{,\alpha}\phi^{,\alpha} + m^2\phi^2)\right). \qquad (19)$$

Decomposing field along the modes (7) and substituting it into (19), for the vacuum expectation we have

$$K \equiv T_{00}^{vac} = \langle 0|T_{00}|0\rangle = \frac{1}{8\pi}\sum_k \left\{ \left|\dot{u}_k\right|^2 + m^2\left|u_k\right|^2 + e^{-2\alpha(t)-2\beta_i(t)}\left|u_{k,i}\right|^2 \right\}. \quad (20)$$

Taking basic modes

$$u_k(t) = \frac{1}{(2\pi)^{n/2}} e^{-n\alpha/2} e^{-i\vec{k}\vec{x}} \chi_k(t),$$

and substituting it into (20), we get

$$K = -\frac{e^{-n\alpha(t)}}{(2\pi)^n 8\pi} \sum_k \left\{ \left|\dot{\chi} - \frac{n}{2}\dot{\alpha}\chi\right|^2 + (m^2 + k_i^2 e^{-2\alpha(t)-2\beta_i(t)})|\chi|^2 \right\}.$$

Separating the term with anisotropy, where's appearing just functions $\beta_i(t)$, for long times we have

$$\Delta K_{anis} = -\frac{e^{-n\alpha(t)-2\lambda t/n}}{8\pi(2\pi)^n} \sum_k F_k |\chi_k|^2, \quad (21)$$

$$F_k = A^i k_i^2.$$

We would use formal WKB-solution for the field $\chi_k(t)$:

$$\chi_k(t) = \frac{1}{\sqrt{2W_k}} \exp\left(-i\int W_k(t')dt'\right),$$

$$W_k^2(t) = \omega_k^2(t) - \frac{1}{2}\left(\frac{\ddot{W}_k}{W_k} - \frac{3}{2}\frac{\dot{W}_k^2}{W_k^2}\right),$$

where

$$\omega_k^2(t) = m^2 - \frac{\lambda^2}{4} + \sum_i k_i^2 e^{-2\alpha(t)-2\beta_i(t)} + \frac{nB^2}{4(n-1)} e^{-2n\alpha(t)},$$

see eq.(13). For the zero's approximation, taking «out» state with

$$\omega_0^2 = m_{eff}^2 = m^2 - \frac{\lambda^2}{4},$$

we finally get

$$\Delta K_{anis} = -\frac{\sqrt{2\Lambda}}{B} \frac{e^{-(1+2/n)\lambda t}}{4(2\pi)^{n+1}\omega_0} \sum F_k, \quad (22)$$

where we've made regularization by offcut for momentum. Because of the divergency we don't have any meaning to get more detailed expression for $\Delta K_{anis}$.

In case of initially massless scalar field it gets the effective mass, as following from eq. (9),

$$m_{eff}^2 = -\frac{\lambda^2}{4} = -\frac{\Lambda n}{2(n-1)},$$

when $\Lambda < 0$.

Of course is interested effect of influence of arising from vacuum particles on expanding and anisotropy of multidimensional space. It needs additional calculations. In case of the model with three space dimensions effect of particle birth leads to fast isotropization [14]. The same could probably be expected also in the high-dimensional theory with metric type (1).

**References:**

1. An Alternative to Compactification. Lisa Randall, Raman Sundrum, Phys.Rev.Lett. 83, 4690(1999) [hep-th/9906064].
2. Large and infinite extra dimensions. V.A.Rubakov, Phys.Usp.44:871-893, 2001 [hep-ph/0104152].
3. U(1) gauge field of the Kaluza-Klein theory in the presence of branes. Gundwon Kang, Y.S. Myung, Phys.Rev.D63:064036, 2001 [hep-th/0007197].
4. Little Randall-Sundrum model and a multiply warped spacetime. Kristian L.McDonald. [hep-th/0804.0654v2], 10 Apr.2008.
5. Structure of space-time. R.Penrose, New York -Amsterdam, 1968.
6. Gravitation. C.W. Misner, K.S. Thorne, J.A. Wheeler, Moscow, 1996.
7. Introduction to the theory of quantum fields. N.N.Bogolubov, D.V.Shirkov. Moscow, 1976.
8. Quantum fields in curved space-time. N.D.Birrell, P.C.W. Davies. Moscow, 1998
9. Introduction to the quantum statistical mechanic. N.N.Bogolubov, N.N.Bogolubov (2), Moscow, 1984.
10. Space-time in the micro-world. D.I.Blohintsev. Moscow, 1982.
11. Quantum effects in intensive external fields. A.A. Grib, S.G.Mamaev, V.M.Mostepanenko. Moscow, 1980.
12. Physical fields in general theory of relativity. N.V. Mitskevich. Moscow, 1969.
13. Methods of mathematical physics. H.Jeffreys, B.Swirles. Cambridge, 1966.
14. About evaluation of radius of compactification and vacuum energy in the Randall-Sundrum model. S.V.Yakovlev, [gr-qc] arXiv:1109.3768v1 [qr-qc], 2011.
15. Investigation of anisotropic metric in multidimensional space-time and generation of additional dimensions by scalar field. S.V.Yakovlev, arXiv: 1109.3769v1 [qr-qc], 2011.